\documentclass{PoS}

\title{Equation of State of Nucleon Matters from Lattice QCD Simulations}

\ShortTitle{Equation of State of Nucleon Matters from Lattice QCD Simulations}

\author{\speaker{Takashi Inoue}\\
        Nihon University, College of Bioresource Sciences\\
        E-mail: \email{inoue.takashi@nihon-u.ac.jp}}

\author{for HAL QCD Collaboration\\
}

\abstract{
Nucleon matters are studied based on QCD. 
We extract nucleon-nucleon interaction from lattice QCD simulations in a recently developed approach,
and then derive the equations of state of the symmetric nuclear matter and the pure neutron matter, 
at zero temperature, in the Brueckner-Hartree-Fock framework. 
We find that QCD reproduce known features of the symmetric nuclear matter,
such as the self-binding and saturation, at some values of quark mass.
We find also that the pure neutron matter become more stiff at large density as quark mass decrease.
We apply these equations of state to neutron star and study its mass and radius.
}

\FullConference{31st International Symposium on Lattice Field Theory - LATTICE 2013\\
		July 29 - August 3, 2013\\
		Mainz, Germany}

\begin{document}

\section{Introduction}
Equation of state (EOS) of dense baryonic matter is a key ingredient for many interesting physics
including heavy ion collision, supernova, and neutron star.
In fact, theoretical prediction of mass and radius of neutron star strongly depend on the EOS.
Therefore, recent discoveries of massive pulsar put severe constraint on the EOS.
While, heavy-ion-collision experiment are revealing EOS of the symmetric nuclear matter.

{\it Ab initio} derivation of the EOS starting from an underlying nuclear force has been
a great challenge and several frameworks are developed, for instance,
the reaction matrix approach~\cite{Brueckner:1958zz},
the cluster variational method~\cite{Pandharipande:1971up},
the relativistic mean field theory~\cite{Walecka:1974qa},
and the quantum Monte Calro simulation~\cite{Gandolfi:2006vp}.
These approaches succeed in explaining known features of nucleon matters
and their predictions have been applied to physics of neutron star.
However, most of microscopic derivation start from a phenomenological nuclear force,
and connection to QCD is rarely considered.
In this paper, we derive nucleon matter EOS starting from QCD.

\section{Nuclear force from QCD}
A new method has been recently proposed to extract nucleon-nucleon ($N\!N$) interaction 
from QCD on the lattice~\cite{Ishii:2006ec} and  already applied to others systems~\cite{Nemura:2008sp}. 
In the method, we utilize
the equal time Nambu-Bethe-Salpeter (NBS) wave function,
defined for the two-nucleon case by
\begin{equation}
 \varphi_{\vec{k}}(\vec r)  = \sum_{\vec x} 
   \langle 0 \vert N(\vec x + \vec r,0)N(\vec x,0)\vert  {N} {N}, \vec{k} \rangle
\label{eqn:psi}
\end{equation}
where $\vert N N, \vec{k}\rangle$ is a two-nucleon QCD eigenstate in the rest frame with a relative momentum $\vec{k}$ and $N(\vec x,t)$ is the nucleon field operator.
%This wave function can be obtained by measuring four point function in lattice QCD numerical simulation,
%and generally contains not only the ground state but also excited states.
From the NBS wave function, we can define a non-local potential $U(\vec r, \vec r')$
though a Schr\"{o}dinger equation as
\begin{equation}
     H_0 \, \varphi_{\vec{k}}(\vec r) 
   + \int \!\! d^3 \vec r' \, U(\vec r, \vec r') \, \varphi_{\vec{k}}(\vec r')
   = E_{\vec k}\varphi_{\vec{k}}(\vec r)
\end{equation}
where $H_0 =  - \frac{\nabla^2}{2\mu}$ and $E_{\vec k}=\frac{\vec{k}^2}{2\mu}$ with the reduced mass $\mu=M_{N}/2$.
Note that the potential $U(\vec r, \vec r')$ can be made energy($\vec k$) independent\cite{Ishii:2006ec}.

It is shown in Ref.~\cite{Inoue:2010es} that the non-local potential  can be efficiently extracted from the 4-pt correlation function in lattice QCD defined by
\begin{eqnarray}
\Psi(\vec{r}, t) &\equiv & \sum_{\vec{x}}\langle 0 \vert N(\vec x + \vec r,t)N(\vec x,t) {\cal J}(t_0)\vert 0 \rangle
=\sum_{\vec{k}} A_{\vec{k}}\varphi_{\vec{k}}(\vec{r})e^{-W_{\vec{k}}(t-t_0)} +\cdots 
\end{eqnarray}
where ${\cal J}(t_0)$ is a source operator which creates two-nucleon states at $t_0$,
the normalization  $A_{\vec k}=\langle NN,\vec{k}\vert {\cal J}(0)\vert0\rangle$,
the total energy $W_{\vec{k}}=2\sqrt{M_N^2+\vec{k}^2}$, and ellipsis denotes inelastic contributions,
which can be ignored for reasonably large $t-t_0$.  
Using the non-relativistic approximation that $W_{\vec k} \simeq 2 M_N + E_{\vec k}$ and employing 
the velocity (derivative) expansion of the non-local potential that
$U(\vec r,\vec r') = \delta^{3}(\vec r-\vec r')V(\vec r,\nabla)
= \delta^{3}(\vec r-\vec r')\left\{ V_0(\vec r) + O(\nabla) \right\}$,
the leading order potential can be extracted as
\begin{equation}
  V_0(\vec r) = \frac{1}{2\mu}\frac{\nabla^2 \Psi(\vec r, t)}{\Psi(\vec r, t)} - 
              \frac{\frac{\partial}{\partial t} \Psi(\vec r, t)}{\Psi(\vec r, t)} - 2 M_N .
\label{eq:vr}
\end{equation}

It is important to note that this extraction does NOT require the ground state saturation for $\Psi(\vec r, t)$, which is usually very difficult or almost impossible
to achieve in actual lattice QCD numerical simulations, in particular on a large spacial volume for two-baryon systems.
Indeed potentials are almost  independent on  $t-t_0$ in this method,
as long as $t-t_0$ is large enough so that  a single hadron propagator is saturated by its ground state and the higher order terms in the derivative expansion are negligible. 
It is also remarkable that the potential is independent on the lattice volume, 
if the volume is larger than the largest interaction range between  hadrons.
In other words, we do NOT need difficult infinite-volume extrapolations in the potential method.
This is a significant advantage  over the conventional method.

%Once we obtain the potential $V_0(\vec r) + \cdots$, we can study any observable of any nuclear system,
%by solving the ordinary Schr\"{o}dinger equation with the potential in the infinite volume.

\section{Nucleon matters from nuclear force}
In order to investigate the nucleon matters, we have to deal with the ground state of interacting infinite nucleon system.
One successful approach is the Brueckner-Bethe-Goldstone (BBG) expansion,
where perturbative expansion is rearranged in terms of the reaction matrix,
and terms are ordered according to number of independent hole-line appearing in its diagrammatic representation.
The lowest two-hole-line approximation with a self-consistent single particle potential,
is called the Brueckner-Hartree-Fock (BHF) framework. We adopt this framework in this paper.

In the BHF framework, total energy of the ground state, $E_0$,
of a nucleon matter at zero temperatures, for nucleon mass $M_N$ and the Fermi momentum $k_F$,
is calculated by
\begin{equation}
 E_0 = \sum_{k}^{k_F} \frac{k^2}{2 M_N} 
     + \frac{1}{2} \sum_{k,k'}^{k_F} \mbox{Re} \langle k k'| G(e(k)+ e(k')) |k k' \rangle_A
\end{equation}
where $G$ is the reaction matrix describing the scattering of two nucleons above the Fermi sphere
and $|k k'\rangle_A = |k k'\rangle - |k' k \rangle$.
This matrix is given by a summation of ladder diagrams representing repeated action of the bare $N\!N$ interaction $V$, 
and is obtained by solving the integral equation
\begin{equation}
   \langle k_1 k_2 |G(\omega)|  k_3 k_4 \rangle
 = \langle k_1 k_2|V|k_3 k_4 \rangle \nonumber 
 + \sum_{k_5, k_6}
    \frac{\langle k_1 k_2|V|k_5 k_6 \rangle \,Q(k_5,k_6) \, 
         \langle k_5 k_6 | G(\omega) |  k_3 k_4 \rangle}
         {\omega - e(k_5) - e(k_6)}
\label{eqn:gmateq}
\end{equation}
where $Q(k,k')=\theta(k-k_F)\theta(k'-k_F)$ is the Pauli exclusion operator
to prevent two nucleons from scattering into occupied states.
The single particle spectrum $e(k) = \frac{k^2}{2M_N} + U(k)$ contains the single particle potential $U(k)$,
which is essential for faster convergence of the BBG expansion.
In the lowest order BHF framework, $U(k)$ is determined from the consistency condition
\begin{equation}
 U(k) = \sum_{k' \le k_F} \mbox{Re} \langle k k'| G(e(k)+ e(k')) |k k'\rangle_A 
\label{eqn:self}
\end{equation}
for both $k \le k_F$ and $k>k_F$ in the continuous choice.

The $G$-matrix is decomposed in the partial waves by using the angle averaged $Q$-operator.
The integral equation for the $G$-matrix (\ref{eqn:gmateq}) 
can be solved easily in the matrix inversion method~\cite{Haftel}. 
First, the self-consistent potential $U(k)$ is determined in the iteration procedure, 
then the ground state energy $E_0$ is calculated. 

\section{Setup of lattice QCD numerical simulation}

For lattice QCD numerical simulations with dynamical quarks,
we use five gauge configuration ensembles generated by using
the renormalization group improved Iwasaki gauge action
and the non-perturbatively $O(a)$ improved Wilson quark action~\cite{Iwasaki:2011np}. 
The setup of simulation parameters are summarized in Table~\ref{tbl:lattice}.
These ensembles are generated to study the flavor $SU(3)$ symmetric world,
and hence the strange quark mass are set equal to up and down quark mass.
The values of quark hopping parameter $\kappa_{uds}$, together with measured hadron masses,
are given in Table~\ref{tbl:mass}.
As you see, with these ensembles, we can study nuclear system at wide range of quark mass.

\begin{table}[t]
\caption{Lattice parameters such as
the lattice size, the inverse coupling constant $\beta$, the clover coefficient $c_{\rm sw}$,
the lattice spacing $a$ and the physical extension $L$. See ref.~\cite{CPPACS-JLQCD} for details.}
\label{tbl:lattice}
\smallskip
\centering
 \begin{tabular}{c|c|c|c|c}
   \hline   \hline
    size             & ~~ $\beta$ ~~ & ~~ $c_{\rm sw}$ ~~ & ~ $a$ [fm] ~ & ~$L$ [fm]~  \\
   \hline 
    $32^3 \times 32$  &   1.83  &   1.761   &  0.121(2)  & 3.87  \\
   \hline  \hline
 \end{tabular}
\end{table}

\begin{table}[t]
\caption{Quark hopping parameter $\kappa_{uds}$ and corresponding hadron masses,
$M_{\rm PS}$, $M_{\rm Vec}$, $M_{\rm Bar}$
for pseudoscalar meson, vector meson and octet baryon, respectively.}
\label{tbl:mass}
\smallskip
\centering
 \begin{tabular}{c|c|c|c|c}
   \hline  \hline
    ~~ $\kappa_{uds}$ ~~ & $M_{\rm PS}$ [MeV] &  $M_{\rm Vec}$ [MeV] &  $M_{\rm Bar}$ [MeV] &
   $N_{\rm cfg}\,/\,N_{\rm traj}$ \\
   \hline 
     0.13660 &   1170.9(7) &   1510.4(0.9) & 2274(2) & 420\,/\,4200 \\
     0.13710 &   1015.2(6) &   1360.6(1.1) & 2031(2) & 360\,/\,3600 \\
     0.13760 & ~\,836.5(5) &   1188.9(0.9) & 1749(1) & 480\,/\,4800 \\
     0.13800 & ~\,672.3(6) &   1027.6(1.0) & 1484(2) & 360\,/\,3600 \\
     0.13840 & ~\,468.6(7) & ~\,829.2(1.5) & 1161(2) & 720\,/\,3600 \\
   \hline  \hline
 \end{tabular}
\end{table}

We measure the nucleon two-point and four-point correlation functions
with the wall source and the Dirichlet boundary condition in the temporal direction.
To enhance the signal over noise, 16 measurements are made for each configuration,
together with the average over forward and backward propagations in time.
Statistical errors are estimated in the Jackknife method.

\begin{figure}[t]
\centering
\includegraphics[width=0.425\textwidth]{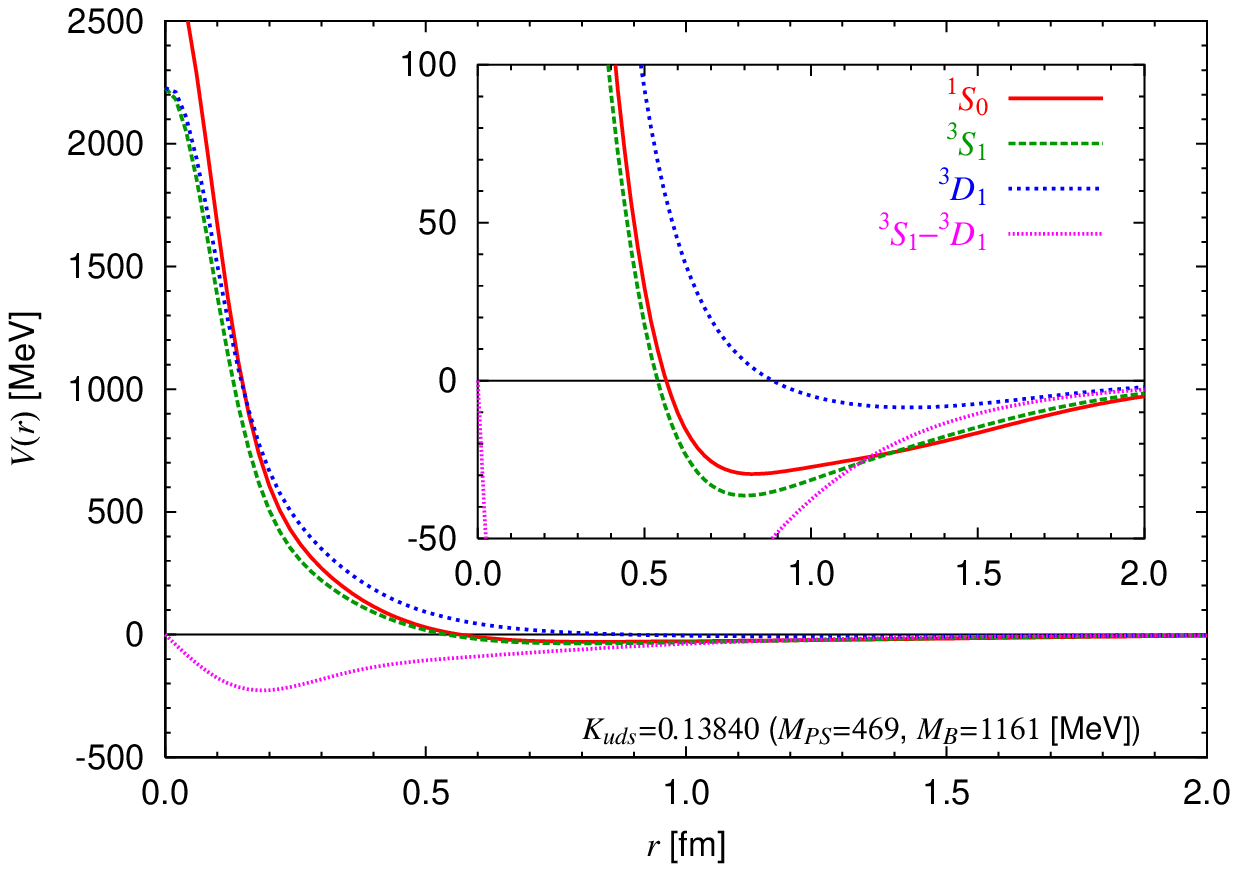}
\quad
\includegraphics[width=0.425\textwidth]{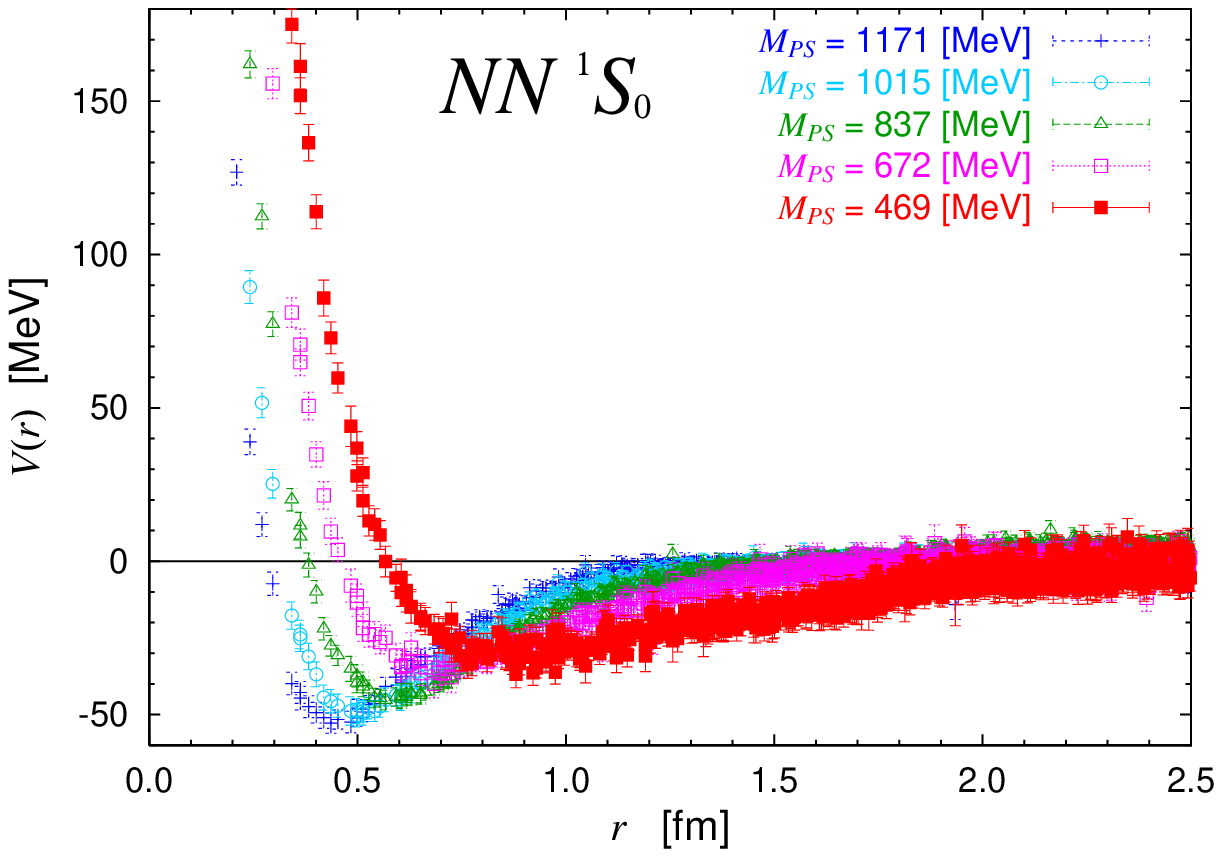}
\caption{Left: Potentials of the S-wave nucleon-nucleon interaction extracted from QCD
at a flavor SU(3) limit corresponding to a pseudoscalar meson mass of 469 MeV.
Analytic functions fitted to the data are used.
Right: Quark mass dependence of $N\!N$ $^1S_0$ potential from QCD.}
\label{fig:pot_K13840}
\end{figure}

Fig.~\ref{fig:pot_K13840}(Left) shows potentials of $N\!N$ interaction in S-wave, 
extracted from a lattice QCD simulation at $\kappa_{uds}=0.13840$,
{\it i.e.} with the present lightest quark corresponding to pseudoscalar meson mass of 469 MeV. 
There, analytic functions fitted to data are used, which are needed to evaluate their matrix element.
The QCD induced $N\!N$ potentials share common features with the phenomenological one (eg.~ref.~\cite{Wiringa:1994wb}),
namely, a repulsive core, an attractive pocket and a strong $^3S_1$-$^3D_1$ coupling.
Therefore, these lattice QCD potentials well reproduce aspect of $N\!N$ scattering observables~\cite{Inoue:2010es}. 
While, strength of the lattice QCD nuclear force is a little weaker than that of empirical one,
especially, the deuteron, {\it i.e} the bound state in $^3S_1$-$^3D_1$ channel, is not supported.
This is because of the heavy $u$, $d$ quark used in the present lattice QCD simulations compared to the physical ones.
When the $N\!N$ potentials are extracted from lattice QCD simulation at the physical point, in near future,
any two-nucleon observables will be reproduced quantitatively. 
Fig.~\ref{fig:pot_K13840}(Right) shows quark mass dependence of $N\!N$ $^1S_0$ potential from QCD. 

\section{Nuclear matter EOS from QCD}

We investigate nucleon matters based on QCD in the BHF framework. 
Namely, we calculate the ground state energy $E_0$ for each of the five ensemble,
by using the lattice QCD extracted $N\!N$ potentials $V(r)$ and the measured value of nucleon mass $M_N$ given in Table~\ref{tbl:mass}.
We truncate the partial wave decomposition of $G$-matrix and include only $^1S_0$, $^3S_1$ and $^3D_1$ channels,
because lattice QCD nuclear force is not available in higher partial waves.

\begin{figure}[t]
\centering
\includegraphics[width=0.375\textwidth]{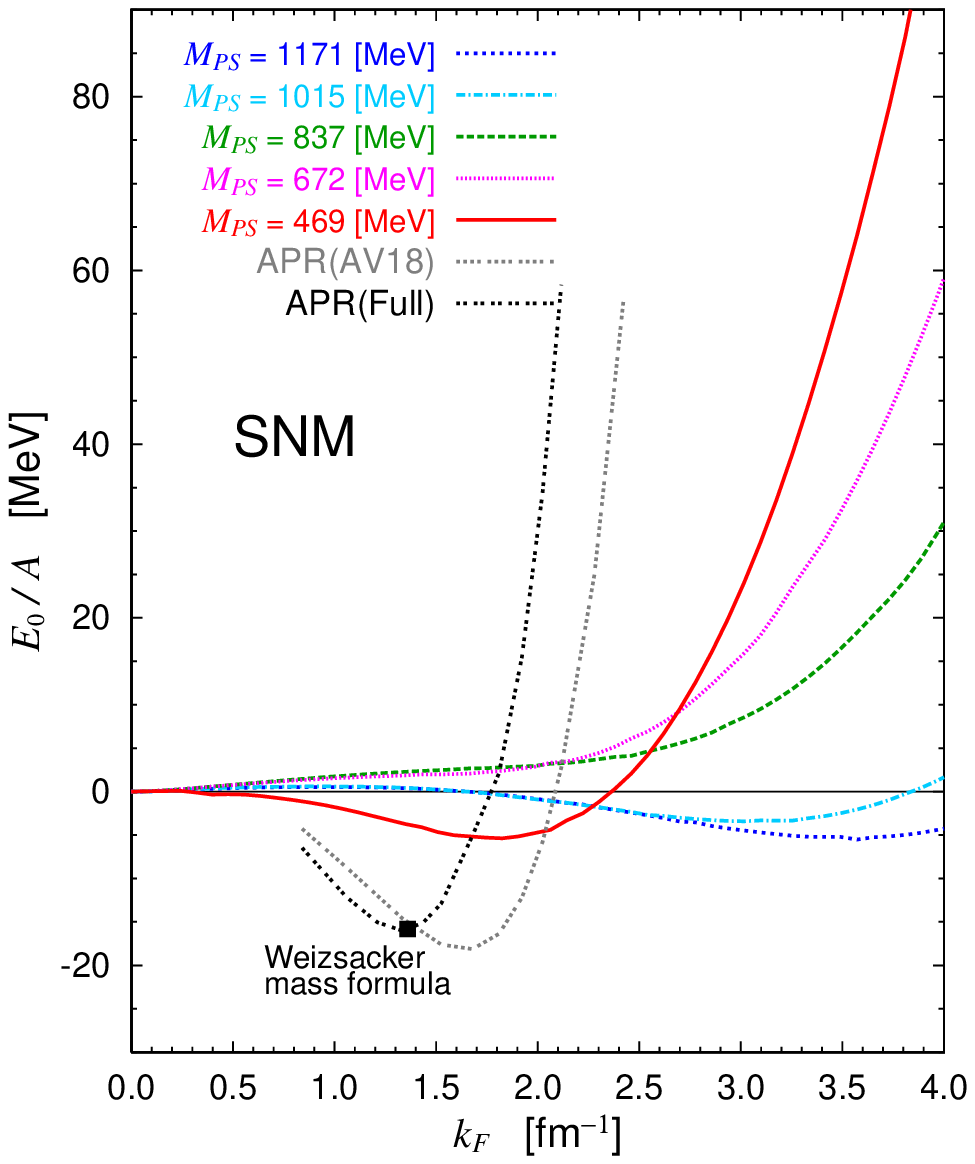}
\quad
\includegraphics[width=0.375\textwidth]{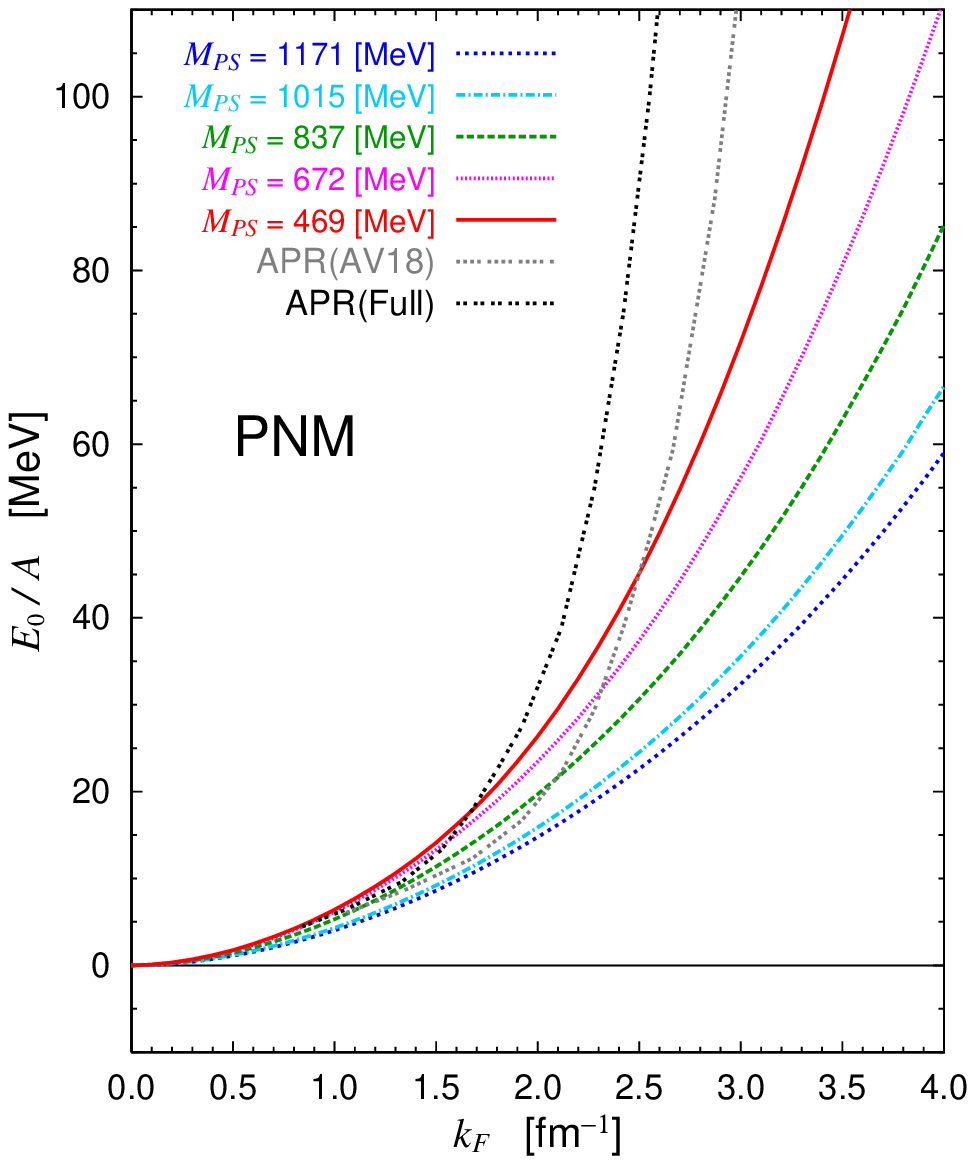}
\caption{Ground state energy par nucleon $E_0/A$ for the symmetric nuclear matter (Left)
and the pure neutron matter (Right), as a function of the Fermi momentum $k_F$.  
The empirical saturation point is also indicated.
The curve labeled APR is taken from ref.~\cite{Akmal:1998cf}}
\label{fig:eos}
\end{figure}

Fig.~\ref{fig:eos}(Left) shows the obtained ground state energy par nucleon $E_{0}/A$ 
for the symmetric nuclear matter (SNM) as a function of the Fermi momentum $k_F$.
The most important feature of the hypothetical matter is the saturation,
where both the binding energy par nucleon and the nucleon density are constant independent on the number of nucleon $A$. 
The empirical saturation point, from the Weizs\"acker mass formula,
is around $(k_F, E_{0}/A) = (1.36~\mbox{fm}^{-1}, -15.7~\mbox{MeV})$, and is also indicated in the figure.
In addition, the curves taken form ref.~\cite{Akmal:1998cf} are shown for a reference, 
which are obtained in the variational method with a modern phenomenological $N\!N$ and $N\!N\!N$ force.

The curve for the lightest quark corresponding to $M_{\rm PS}=469$ MeV, clearly shows the saturation.
This is a significant success, although the obtained saturation point deviate from the empirical one.
It should be noted that we have never used any phenomenological inputs for the $N\!N$ interaction, but used only QCD.
Again, a main reason of the deviation is the unphysically heavy $u$, $d$ quark used in our lattice QCD simulation.
Because the lattice QCD $N\!N$ interactions are weaker than the real ones, 
resulting binding energy is smaller than the empirical value.
From the quark mass dependence of the curve shown in the figure, 
one can expect that a better agreement to data will be achieved,
when we have extracted $N\!N$ interactions from QCD at the physical point.
We see also that the saturation feature is so delicate against change of quark mass,
that it is lost even for the second lightest quark in this study.
It seems that the saturation appears again in the world where quarks are very heavy. 

Fig.~\ref{fig:eos}(Right) shows the obtained ground state energy par nucleon $E_{0}/A$
for the pure neutron matter (PNM) as a function of $k_F$.
Since the pressure of matter is given by a slope of the curve, %$P \propto \frac{d \, (E_0/A)}{d k_F}$,
these curves are nothing but EOS of the pure neutron matter at zero temperature.
We can see that, at a given large density, the pressure is larger for the matter with lighter quark.
This result shows that the pure neutron matter become more stiff as quark mass decrease.
Again, the lattice QCD prediction seems to approach to phenomenological one as quark mass decrease.

\section{Discussion}

To see effect of the stiffening of nucleon matters, let us study mass and radius of neutron star. 
We solve the Tolman-Oppenheimer-Volkoff equation~\cite{Oppenheimer:1939ne}
with the EOS of neutron-star matter,
which consists of neutrons, protons, electrons and muons under the charge neutrality and beta equilibrium.
For EOS of arbitrary asymmetric nuclear matter, we use the parabolic approximation,
${E_0}/{A}(\rho,x) = {E_0^{\rm SNM}}/{A}(\rho) + (1 - 2 x)^2 E_{\rm sym}(\rho)$,
which is the standard interpolation between SNM and PNM 
with the symmetry energy $E_{\rm sym}(\rho)$ and the proton fraction $x = \rho_p/\rho$.
The symmetry energy is defined by $E_{\rm sym}(\rho) = E_0^{\rm PNM}/A(\rho) - E_0^{\rm SNM}/A(\rho)$.
%with the energy of PNM and SNM at the same density $\rho$.
Electrons and muons are treated as the non-interacting Fermi gas. 
We ignore the crust part of neutron star.

\begin{figure}[t]
\centering
\includegraphics[width=0.4\textwidth]{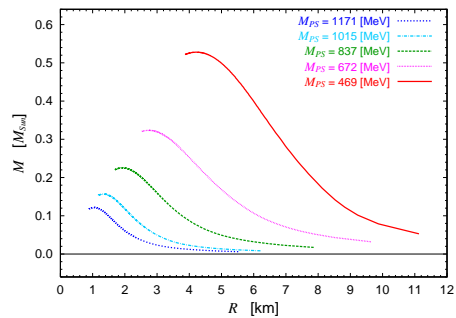}
\caption{Mass-radius relation of neutron star derived from the lattice QCD nuclear force.}
\label{fig:neutron_star}
\end{figure}

Fig.~\ref{fig:neutron_star} shows the mass-radius relation of stable and non-rotating neutron stars,
derived from lattice QCD nuclear force at the five quark masses.
We can see non-trivial dependence of neutron stars on quark mass.
The maximum mass of neutron star ranges from 0.12 to 0.53 $M_{\odot}$ with the solar mass $M_{\odot}$. 
These values are much smaller than the mass of already observed neutron stars.
Of course, this failure is due to the unphysically heavy $u$, $d$ quark in our simulations.
Since the maximum mass increase rapidly as quark mass decrease,
one can fairly expect that QCD with physical quarks
will predict more reasonable value consistent to astrophysical observations.

In this paper, we have studied the nucleon matters and neutron stars starting from QCD, 
the fundamental theory of strong interaction.
We have included S-wave $N\!N$ interactions mainly so far. 
We have plan to include $P$-wave and $D$-wave $N\!N$ interactions~\cite{Murano:2013xxa}
and three-nucleon ($N\!N\!N$) force~\cite{Doi:2011gq} in this study.
It is expected that hyperons should exist in the inner core of neutron star.
To study the hyperon onset theoretically, hyperon-nucleon ($Y\!N$) and hyperon-hyperon ($Y\!Y$) interactions are needed.
Today, these interactions are not well revealed from experiment unfortunately.
While, from lattice QCD, we can extract both the $Y\!N$ and $Y\!Y$ interactions 
as well as the $N\!N$ interaction~\cite{Nemura:2008sp,Inoue:2010es}.
Therefore, it is interesting and important to investigate the hyperon onset in neutron star matter by basing on QCD.
A study toward this target is in progress.

\begin{acknowledgments}
We thank K.-I. Ishikawa and PACS-CS group for providing their DDHMC/PHMC code~\cite{Aoki:2008sm},
and authors and maintainer of CPS++~\cite{CPS}, whose modified version is used in this study.
Numerical computations of this work have been carried out at Univ.~of Tsukuba supercomputer system (T2K).
This research is supported in part by Grant-in-Aid for Scientific Research (C)23540321.
\end{acknowledgments}

\end{document}